\documentclass{nature}
\usepackage[left=1.8cm,top=2cm,right=1.8cm,bottom=1.8cm,nohead]{geometry}

\def\arcsec{''}
\def\asec{\ifmmode ''\!. \else $''\!.$\fi}
\newcommand\fs{\hbox{$.\!\!^{s}$}}

\def\micron{\ifmmode \mu{\rm m} \else $\mu$m\fi}

\def\Msun{M$_{\odot}$}

\def\HST{{\it HST}}
\def\sst{{\it Spitzer}} 
\def\Y{{F105W}}

\def\JH{{F140W}}
\def\H{{F160W}}
\def\apj{{\it Astrophys. J.}}
\def\apjs{{\it Astrophys. J. Suppl.}}
\def\aj{{\it Astron. J.}}
\def\aap{{\it A\&Ap}}
\def\mnras{{\it Mon. Not. R. Astron. Soc.}}
\def\cl{MACS J1149.6+2223}
\def\obj{MACS1149-JD1}
\newcommand {\lya}{Ly$\alpha$}

\newcommand{\gsim}{\mbox{\hspace{.2em}\raisebox{.5ex}{$>$}\hspace{-.8em}\raisebox{-.5ex}{\
$\sim$}\hspace{.2em}}}
\newcommand{\lsim}{\mbox{\hspace{.2em}\raisebox{.5ex}{$<$}\hspace{-.8em}\raisebox{-.5ex}{\
$\sim$}\hspace{.2em}}}

\title{A highly magnified candidate for a young galaxy seen when the Universe 
was 500 Myrs old}

\author{
Wei Zheng$^{1}$, 
Marc Postman$^{2}$,
Adi Zitrin$^{3}$,
John Moustakas$^{4}$,
Xinwen Shu$^{5}$,
Stephanie Jouvel$^{6,7}$, 
Ole Host$^{6}$,
Alberto Molino$^{8}$,
Larry Bradley$^{2}$,
Dan Coe$^{2}$,
Leonidas A. Moustakas$^{9}$,
Mauricio Carrasco$^{10}$,
Holland Ford$^{1}$,
Narciso Ben\'itez$^{8}$,
Tod R. Lauer$^{11}$,
Stella Seitz$^{12}$,
Rychard Bouwens$^{13}$,
Anton Koekemoer$^{2}$,
Elinor Medezinski$^{1}$,
Matthias Bartelmann$^{3}$,
Tom Broadhurst$^{14}$,
Megan Donahue$^{15}$,
Claudio Grillo$^{16}$,
Leopoldo Infante$^{10}$,
Saurabh Jha$^{17}$,
Daniel D. Kelson$^{18}$,
Ofer Lahav$^{6}$,
Doron Lemze$^{1}$,
Peter Melchior$^{19}$,
Massimo Meneghetti$^{20}$,
Julian Merten$^{3}$,
Mario Nonino$^{21}$,
Sara Ogaz$^{2}$,
Piero Rosati$^{22}$,
Keiichi Umetsu$^{23}$,
Arjen van der Wel$^{24}$
}
\begin{document}
\maketitle

\begin{affiliations}
\item {Johns Hopkins University, 3701 San Martin Drive, Baltimore, MD 21218, U.S.A.}
\item {Space Telescope Science Institute}
\item {Universit\"at Heidelberg}
\item {University of California, San Diego}
\item {University of Science and Technology of China}
\item {University College London} 
\item {Institute de Ciencies de l'Espai}
\item {Instituto de Astrof\'isica de Andaluc\'ia}
\item {Jet Propulsion Laboratory, California Institute of Technology}
\item {Pontificia Universidad Cat\'olica de Chile}
\item {National Optical Astronomical Observatory}
\item {Universitas Sternwarte, M\"unchen}
\item {Leiden Observatory}
\item {University of Basque Country}
\item {Michigan State University}
\item {Niels Bohr Institute, University of Copenhagen}
\item {Rutgers University}
\item {The Observatories of the Carnegie Institution for Science}
\item {The Ohio State University}
\item {INAF-Osservatorio Astronomico di Bologna}
\item {INAF-Osservatorio Astronomico di Trieste}
\item {European Southern Observatory}
\item {Academia Sinica, Institute of Astronomy \& Astrophysics}
\item {Max-Planck Institut f\"ur Astronomie}
\end{affiliations}

\begin{abstract} 
The early Universe at redshift {\emph z}$\sim$6-11 marks 
the reionization of the intergalactic medium,
following the formation of the first generation of stars.
However, those young galaxies at a cosmic age of $\lsim$500 million years (Myr, 
at {\emph z} $\gsim$ 10) remain largely 
unexplored as they are at or beyond the 
sensitivity limits of current large telescopes.
Gravitational lensing by galaxy clusters enables the detection of 
high-redshift galaxies that are fainter than what otherwise could be 
found in the deepest images of the sky. We report the discovery of an object found in the 
multi-band observations of the cluster MACS1149+22 that has a high probability
of being a gravitationally magnified object from the early universe. 
The object is firmly detected (12\mbox{\boldmath${\sigma}$}) in the two reddest 
bands of \emph{HST}/WFC3,
and not detected below 1.2 \mbox{\boldmath${\mu}$}m, matching the characteristics of 
\emph{z}$\sim$9 objects.
We derive a robust photometric redshift of \emph{z} = 9.6 $\pm$ 0.2, 
corresponding to a cosmic age of 490 $\pm$ 15 Myr (\emph{i.e.}, 
3.6\% of the age of the Universe). The large number of bands used to derive the 
redshift estimate make it one of the most accurate estimates ever obtained for such 
a distant 
object.
The significant magnification by cluster lensing (a factor of $\sim$15) 
allows us to analyze the object's ultra-violet and 
optical luminosity in its rest-frame, thus enabling us to constrain on
its stellar mass, star-formation rate and age.
If the galaxy is indeed at such a large redshift, then its age 
is less than 200 Myr (at the 95\% confidence level), implying a formation redshift 
of \emph{z}$_f\lsim$14.
The object is the first \emph{z}$>$9 candidate that is bright enough for detailed 
spectroscopic studies with \emph{JWST},
demonstrating the unique potential of galaxy cluster fields for finding highly 
magnified, intrinsically faint galaxies at the highest redshifts.
\end{abstract}

Observational cosmology has established that the age of the Universe is 13.7 
billion years, and the reionization of the vast intergalactic medium (IGM) 
started around redshift $z \sim 11$,\cite{wmap} as the result of radiation 
from the first generation of stars. 
The task of probing the most distant Universe is progressively challenging: 
While more than $10^5$ quasars have been found, 
only one is at $z > 7$;\cite{ukirt}  
while thousands of gamma-ray burst events have been recorded, only one\cite{grb} 
is confirmed at \emph{z}=8.3;
and while thousands of galaxy candidates have been found at $z \sim 6$, only one 
has been reported at $z\sim 10$,\cite{bouwens} 
which is based on a single-band detection. 
Galaxies at $z \sim 10$ are expected to be at a magnitude of $\sim 29$ 
(in the AB system, used hereafter)\cite{bouwens,oesch}, 
near the detection limits of the deepest fields observed by {\it Hubble Space Telescope} 
(\HST), and beyond the spectroscopic capability of even the next generation of
large telescopes.

In this {\it Letter} we report the discovery of a gravitationally lensed source whose most likely redshift is $z\sim 9.6$.
The source, hereafter called \obj, is selected from a near-infrared detection image at significance of $22 \sigma$. \obj\ 
has a unique flux distribution characterized by a) no detection at wavelengths shorter than 1.2 \micron,
b) firm detections in the two reddest \HST\ bands and c) weak detections in two 
other \HST/WFC3/IR (Wide-Field Camera 3/Infrared Channel) bands and in one {\it Spitzer}/IRAC 
(Infrared Array Camera) channel. 
The object's coordinates (J2000) are:
RA=$11^h 49^m 33\fs 584$ ~ Dec=+$22^\circ 24^\prime 45\asec 78$.

Galaxy clusters are the largest reservoirs of gravitationally bound dark 
matter (DM), whose huge mass 
bends light and forms ``cosmic lenses.'' They can significantly magnify
the brightness and sizes of galaxies far behind them, thereby revealing 
morphological details that are otherwise impossible to 
detect\cite{kneib,bradley,zheng,richard,bradley2,zitrin} 
and enabling spectroscopy to study the physical conditions in these intrinsically 
faint galaxies. This is particularly important 
for the \sst\ {\it Space Telescope} infrared data where the telescope's low 
spatial resolution blends faint sources.  
At $z>7$, the \lya\ break, at $\sim 0.12 (1+z)$~\micron,
is redshifted out off the optical bands, and the Balmer break,
at $\sim 0.38 (1+z)$~\micron, is redshifted into the \sst/IRAC range. By combining the \HST\ 
and \sst\ data we are able to estimate the age of such distant objects based on 
the ratio of their rest-frame ultra-violet to optical fluxes. 
\obj\ is approximately $15\times$ brighter than it would be in an unlensed 
field.  

The Cluster Lensing And Supernova survey with Hubble (CLASH)\cite{postman} 
is a \HST\ Multi-Cycle Treasury program that acquires 
images in 16 broad bands between $0.2-1.7$ $\mu$m for 25 clusters. 
\cl\ is a massive cluster at redshift $z=0.544$, selected from a group of 
X-ray luminous clusters. The 
mass models for this cluster\cite{zitrin1,smith} suggest a relatively flat mass 
distribution profile and a large area of high magnification,
making it one of the most powerful cosmic lenses known. 

The spectral-energy distribution (SED) features of galaxies, most notably the Lyman break
and the Balmer break,
generate distinct colors between broad bands and enable us to derive their redshifts with 
reasonable accuracy.
Our photometric redshift estimates are made with two different techniques: 
Le Phare (LPZ)\cite{lpz} and Bayesian Photometric Redshifts (BPZ)\cite{bpz}. 
LPZ photometric redshifts are based on a 
template fitting procedure with a maximum likelihood ($\chi^2$) estimate.
We use the template library 
of the COSMOS survey\cite{cosmos}, including galaxy templates of three ellipticals, 
seven spirals\cite{polletta} and 12 common templates\cite{bc03}, with starburst ages 
ranging from 30 Myr to 3 Gyr (billion year) to better reproduce the bluest galaxies. 
The LPZ solution from the marginalized posterior is $z=9.60 ^{+0.20}_{-0.28}$ 
(at 68\% confidence level), and the best-fit model is a starburst galaxy.

BPZ multiplies the likelihood by the prior probability of a galaxy with an 
apparent magnitude $m_{0}$ of having a redshift $z$ and spectral type $T$.
We run BPZ using a new library composed of 11 SED 
templates originally drawn from PEGASE\cite{fioc} but 
recalibrated using the FIREWORKS photometry and spectroscopic redshifts\cite{wuyts} 
to optimize its performance. This galaxy library 
includes five templates for ellipticals, 
two for spirals, and 
four for starbursts. 
The most likely BPZ solution is 
a starburst galaxy at $z=9.61_{-0.13}^{+0.14}$ ($1\sigma$). 

Even though the CLASH data have more bands than other \HST\ projects, \obj\ is 
detected only in the four reddest \HST\ bands. The high confidence of our high-redshift 
solution is enabled by the IRAC photometry at 3.6\micron\ and 4.5\micron. 
With \HST\ data alone (excluding \sst\ data)
solutions with intermediate redshifts ($2 \lsim z\lsim 6$) can be found but they have 
low probability (Fig.~3). When \sst\ data are included, no viable
solutions other than those at $z \sim 9.6$ are found, and the possibility for photometric redshifts 
$z<8.5$ is rejected at $4 \sigma $ confidence level ($< 3\times 10^{-5}$).

Using confirmed multiply-lensed images, strong-lensing (SL) models\cite{zitrin1,zitrin2} allow us to derive the 
mass distribution of DM in the cluster, which leads to an amplification map
for background sources.
With 23 multiply-lensed images of seven sources, we derive the best-fit model 
in which the critical curve (of high magnification) of $z\sim 10$
extends to the vicinity of \obj, resulting in a magnification factor of $\mu =14.5^{+4.2}_{-1.0}$. 
The results are in rough agreement with a second, independent model\cite{jullo}, which yields a
best-fit magnification with large error bars, $26.6^{+20.8}_{-7.7}$.

Because our data cover a broad range in the object's rest-frame, we are able to
estimate some key properties for the source
using the Bayesian SED-fitting code
{\tt iSEDfit}\cite{moustakas} 
coupled to state-of-the-art population
synthesis models\cite{conroy} and based on the Chabrier\cite{chabrier}
initial mass function from $0.1-100$~\Msun\ (solar mass).  We
consider a wide range of parameterized star formation histories and
stellar metallicities and assume no dust attenuation, as previous
studies\cite{labbe1,bouwens2} found no evidence for dust in galaxies at the
highest redshifts.

Fig.~4 presents the results of our population synthesis modeling
adopting $z = 9.6$ as the source redshift.
Based on the median of the posterior probability distributions, our
analysis suggests a stellar mass of $\sim 1.5
\times10^8~(\mu/15)^{-1}$ \Msun\ and a star-formation rate (SFR) of
$\sim1.2~(\mu/15)^{-1}$~\Msun~yr$^{-1}$.  Given the uncertainties in
the IRAC photometry, we are unable to measure the age of the galaxy
precisely; however, we can constrain its SFR-weighted age, or the age
at which most of the stars formed, to $\langle t\rangle_{\rm
SFR}<200$~Myr ($95\%$ confidence level), suggesting a likely formation
redshift $z_{f}<14.2$.  Given that the source is brighter at
4.5\micron\ than at 3.6\micron, the presence of a Balmer break is likely,
suggesting that \obj\ may not be too young. This age implies
a formation redshift of no earlier than $z_{f}\approx 11.3$, and is
generally consistent with the estimated ages ($\gsim 100$~Myr) of
galaxies at slightly lower redshifts, $z \sim 7-8$.\cite{labbe2}

\obj\ is compact, but spatially resolved. 
We can get a clearer view of the source by removing the effects of \HST's 
point spread function (PSF)
using Lucy-Richardson deconvolution\cite{lucy,richardson}.
After deconvolution with the WFC3/IR PSF, the observed ({\it i.e.}, de-magnified)
half-light radius of the core is $r < 0\asec 13$. 
The expected intrinsic size, based on extrapolations from measured galaxy sizes\cite{oesch2}
from $z \sim 7$ to 10 in unlensed fields, is $r=0\asec 07$.  Assuming symmetrical amplification, 
\obj's magnified size would be $\sim 0\asec 25$ before convolution with the WFC3/IR PSF.  
This implies that the source is approximately half the size as expected.
For comparison, if \obj\ is at $z \sim 3$, its expected intrinsic size would be $0\asec 17$, 
and the expected observed size would be $0\asec 41$.  The larger difference between 
the intermediate-redshift expectation and what we observe supports the high-redshift solution.

\noindent{\bf Methods summary}

The presence of a prominent
\lya\ break caused by IGM absorption is used to identify high-redshift galaxy candidates.
We select $z\sim 10$ candidates with the following criteria: 
(1) The difference in magnitude F110W$-$F140W $> 1.3$; (2) F140W$-$F160W $< 0.5$; 
and (3) No detection in the \Y\ band ($< 2 \sigma$) 
and no detection in the optical detection image ($< 1 \sigma$). \obj\ is the only 
object in our current 
database of 12 CLASH clusters observed to date that meets these criteria.
We measure the photometry with a circular aperture of diameter 10 pixels (0\asec65), and 
apply an aperture correction of $-0.3$ magnitude 
to infer the total galaxy magnitude measured with an aperture of 20 pixels.

We verify that \obj\ is not a solar system object by placing an upper limit
on its proper motion over the course of 80 days. The object is also inconsistent with being 
a late-type Galactic star -- there are no L,M,T, or Y dwarfs whose total flux difference is
within $6.8 \sigma$ of the observed colors of \obj. The likelihood that the source is
at an intermediate redshift is extremely low given the \sst/IRAC photometric constraints.
We also study several pairs of flux ratios in different bands for all the objects 
in our CLASH database with similar magnitudes.
The hypothesis that the source's extremely red color is just due to photometric scatter of the general
faint extragalactic population is excluded at 99.985\% 
confidence level. 
More details are presented in the {\it Supplementary Information} section.

\hskip 1.5in

\clearpage

\begin{addendum}
 \item[Acknowledgments] 
The CLASH program (GO-12065) is based on observations 
made with the NASA/ESA {\it Hubble Space Telescope}. The Space Telescope Science 
Institute is operated by the Association of Universities for  Research in 
Astronomy, Inc. under NASA contract NAS 5-26555. 
This work is also based in part on archival data obtained with the {\it Spitzer Space Telescope}, 
which is operated by the Jet Propulsion Laboratory, 
California Institute of Technology under a contract with NASA.

\item[Author Contributions]
W.Z. made the initial identification and wrote a draft. 
R.B., D.C., H.F. and L.B. verified the target selection. 
M.P and H.F performed comparisons with intermediate-redshift and nearby objects and edited the final version. 
W.Z., A.K., L.B., D.C., S.O. and E.M. processed the \HST\ data.
X.S., W.Z. and L.A.M. performed the IRAC photometry. 
S.J., A.M., D.C., O.H. and N.B. made the redshift estimates.
M.P., T.R.L. and L.B. performed the image deconvolution. 
J.M. carried out the SED fitting.  
A.Z., M.C. and T.B. ran the lensing models.
The above authors also contributed the text and figures that describes their analyses.
P.R., L.I., P.M., M.N., R.B. and L.A.M. contributed to the observing programs.
M.B., M.D., C.G., S.J., D.D.K., O.L., D.L., P.M., K.U. and A.W. 
helped with the manuscript.
\item[Correspondence] Correspondence and requests for materials
should be addressed to W.Z. (email: zheng@pha.jhu.edu).
\end{addendum}

\clearpage

\begin{figure} 
\parbox{5.5 in}{
\includegraphics{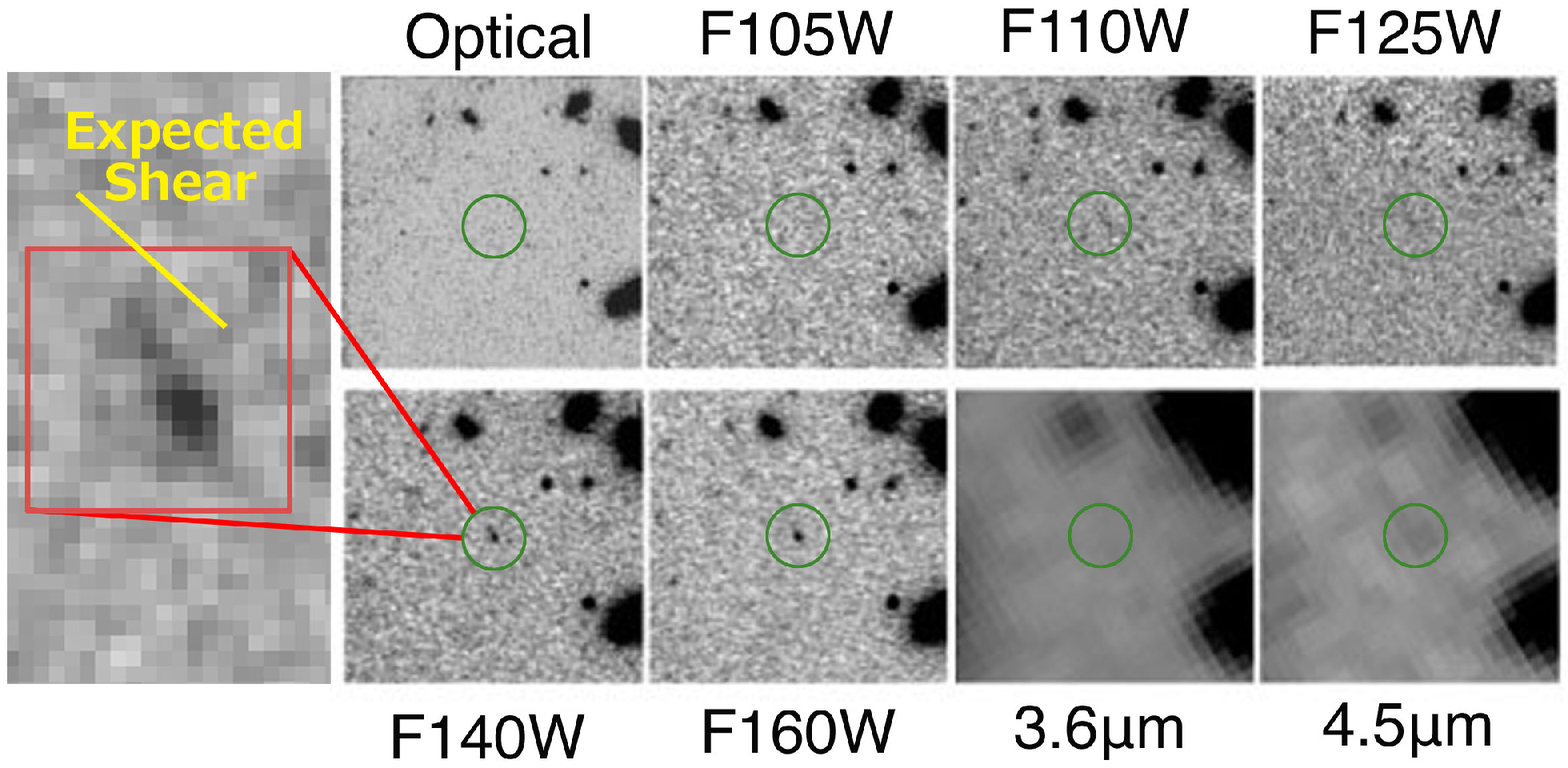} 
}

\vspace{5.95 in}
\noindent {\sf Figure~1 --  
Cutout images of 
\obj\ in the optical (ACS, summed), near-infrared (WFC3) and infrared (IRAC) bands. 
Each image is 10\arcsec\ 
on one side. North is up and east to the left.
The source, located at the center of each image, is firmly detected in
the F140W (1400 nm) and F160W bands and weakly detected in the F110W, F125W and 
4.5\micron\ bands. 
An enlarged view of the \JH\ image shows its elongation, which is
extended along a position angle of $\sim$37 degrees. A yellow line marks the 
direction of shear predicted by the lensing model.
}
\label{stamps}
\end{figure}
\clearpage

\begin{figure} 
\parbox{5.5 in}{
\includegraphics{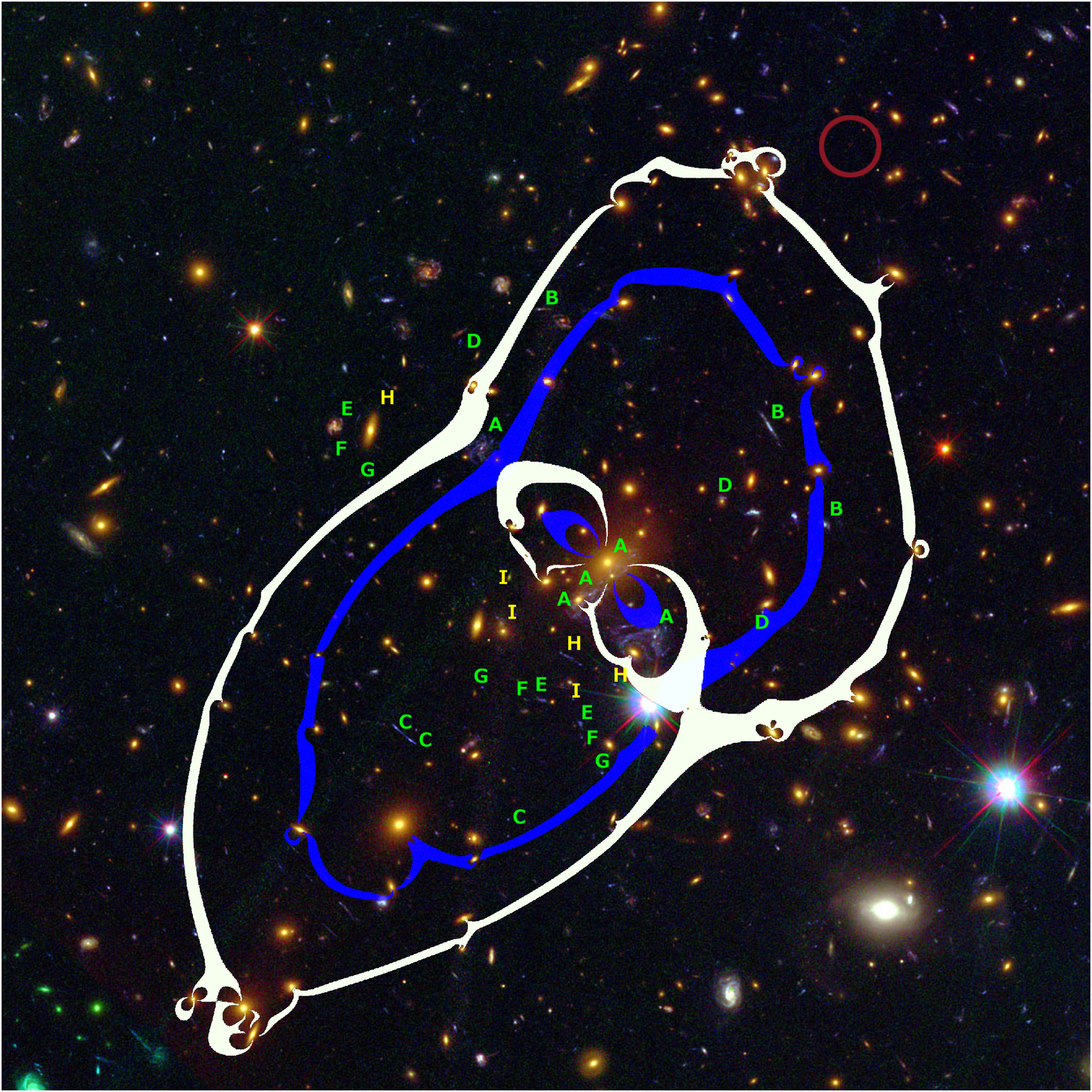} 
}

\vspace{6.0 in}
\noindent {\sf Figure~2 --  Composite color image of \cl. North is up and east to the left. 
The field of view is 2.2 arcmin on each side. The \emph{z} = 9.6 critical curve 
for the best-fit
lensing model is overplotted in white, and
that for \emph{z} = 3 is shown in blue.
Green letters A-G mark the multiple images of seven sources that 
are used in the strong-lensing model. Yellow letters H and I mark the two systems that are not 
used in the 
final fitting. The location of 
\obj\ is marked with a red circle,
at RA=$11^h 49^d 33\fs 584$ ~ Dec=+$22^{\circ} 24^\prime 45\asec 78$ (J2000).
}
\label{color}
\end{figure}
\clearpage

\begin{figure} 
\parbox{5.5 in}{
\includegraphics{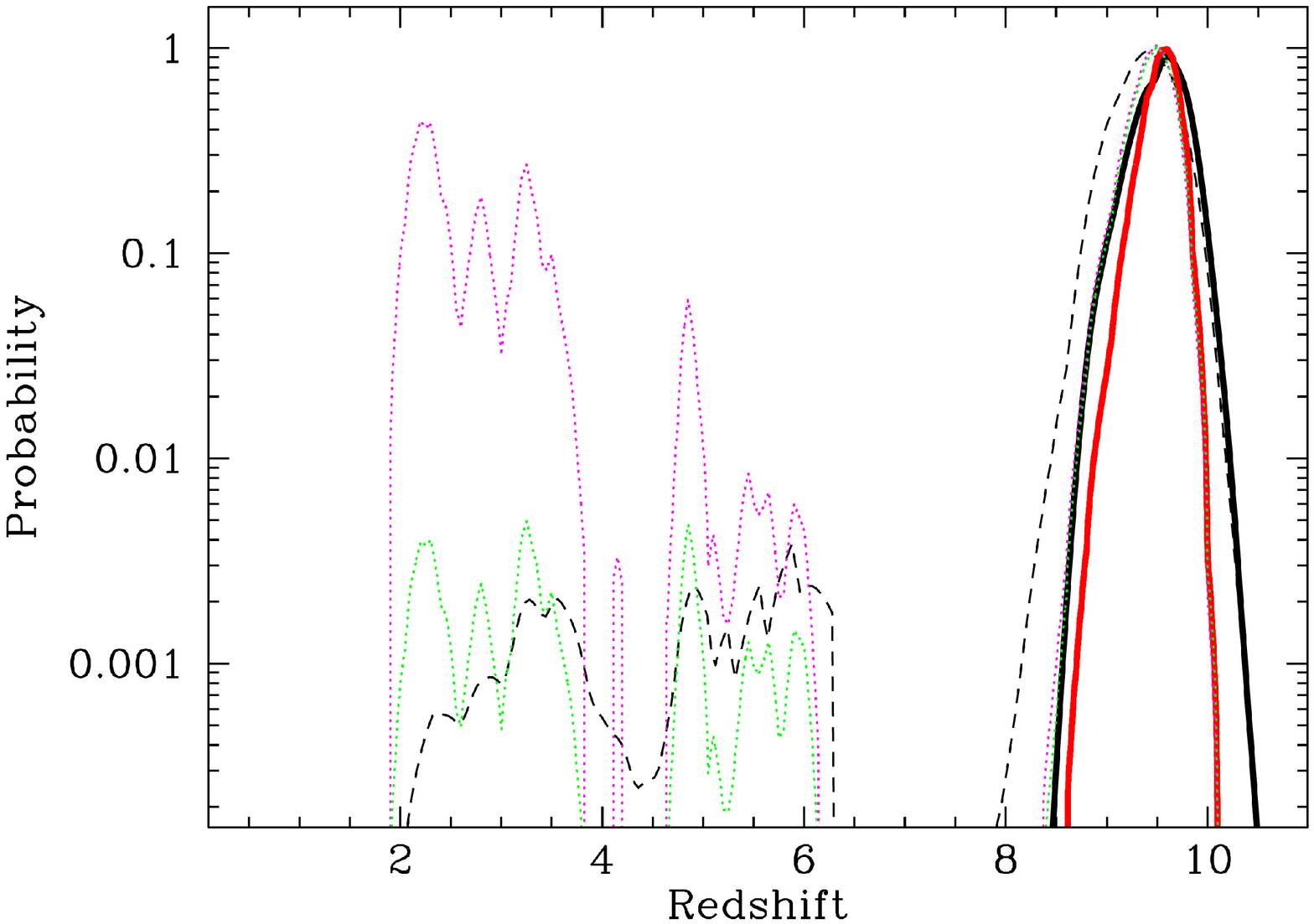} 
}

\vspace{5.95 in}
\noindent {\sf Figure~3 --  
Probability distribution of photometric redshift estimation. 
All curves are normalized to their peak probability. 
Solid black curve: LPZ, using all the \emph{HST} and \emph{Spitzer} data;
Solid red curve: BPZ with and without priors, using all data.
Only the high-redshift solutions are confirmed with high confidence ($> 4 \sigma$).
Dashed black curve: LPZ, using the \emph{HST} data only.
Dotted green curve: BPZ without priors, using the \emph{HST} data only.
In these two cases, intermediate-redshift solutions are present at low probability ($< 1\%$). 
Dotted magenta curve: BPZ with priors, using the \emph{HST} data only. 
Only in this case intermediate-redshift solutions become significant. 
}
\label{pdz}
\end{figure}
\clearpage

\begin{figure} 
\parbox{5.5 in}{
\includegraphics{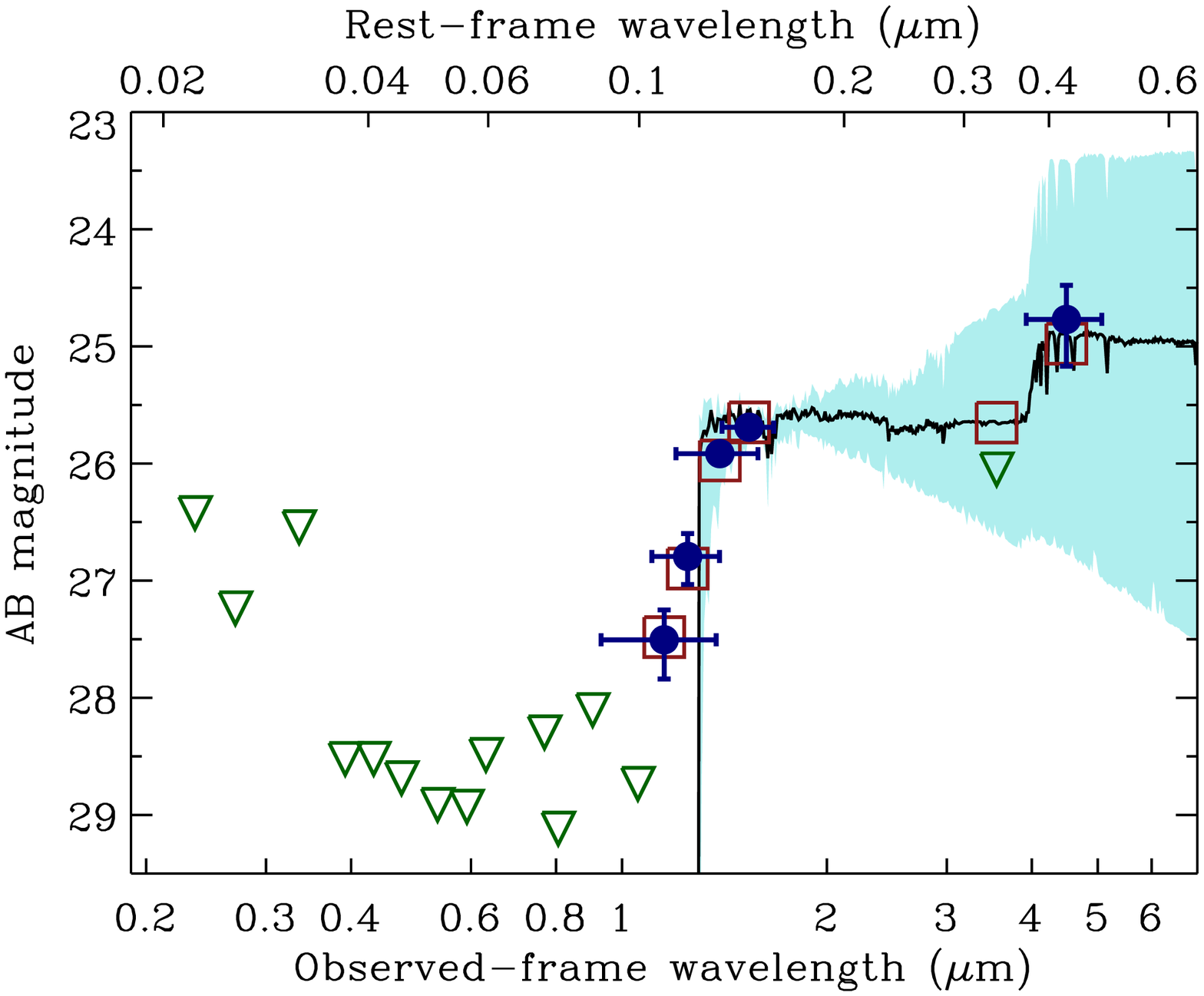} 
}

\vspace{5.95 in}
\noindent {\sf Figure~4 --  
Stellar population synthesis modeling results for \obj.  The filled
blue points mark bands in which the object is detected, while the open
green triangles indicate 1$\sigma$ upper limits. The errors in the F140W and F160W 
bands are small ($<$0.1 magnitude) and hence not visible. The black spectrum
is the best-fit model, and the open red squares show the photometry of
this model convolved with the WFC3, ACS, and IRAC filter response
functions. The light blue shading shows the range of 100 additional
models drawn from the posterior probability distribution that are also
statistically acceptable fits to the data.  
}
\label{sed}
\end{figure} 

\vskip 0.5in

\centerline{\bf \large Supplementary information}
\vskip 0.1in

\centerline{\bf 1.~~  General Outline}

\obj\ is found in a search of all the observations of 12 CLASH clusters. Its characteristics are described 
here in further detail, along with our analysis methods. 
In \S 2 we describe the \HST\ data processing and aperture photometry;
in \S 3 we describe the IRAC photometry that we perform;
in \S 4 we test the intermediate-redshift probability of the source, using only part of the data;
in \S 5 we discuss our lensing model and compare with another independent model; 
in \S 6 we describe the \HST\ image deconvolution; 
in \S 7 we show our SED fitting method; 
in \S 8 we demonstrate why the source is not a solar system or Galactic interloper; 
in \S 9 we discuss the effect of photometric scattering of intermediate-redshift objects; and
in \S 10 we summarize our tests.

We adopt the cosmological parameters
$h = 0.7$, $\Omega_M = 0.3$, and $\Omega_\Lambda = 0.7$. 

\vskip 0.1in

\centerline{\bf 2.~~  \emph{HST} photometry}

The CLASH observations of \cl\ were made between December 2010 and March 2011. 
The \HST\ images include archival data in the F555W and F814W bands. 
The data are processed in two independent pipelines: {\em APLUS}, 
an enhanced version of {\tt APSIS}\cite{apsis} that is now capable of merging and 
aligning WFC3 images, and {\tt Multidrizzle}\cite{mosaic,candels}. They are
combined, aligned and resampled with a common pixel scale of 0\asec065.
A Subaru image, centered on the cluster, but covering a $28 \times 28$ arcminute field is used as the 
astrometric reference.
Detection images are produced from the combination of ACS (Advanced Camera for Surveys)/WFC and WFC3/IR images. 
We use the WFC3/IR detection image and run SExtractor\cite{sex} in dual mode in 
every filter band. 

We carry out photometry with circular apertures whose diameter is
between 2 and 20 pixels. At larger apertures, the source flux in each band increases
and gradually approaches an asymptotic value.
The spectral break becomes less prominent at larger apertures as more noise is 
added to each band. 
At each aperture size we estimate the photometric redshift.
The most precise photometric redshift is from the photometry made
with an aperture of diameter 10 pixels (0\asec65). To verify the precision, we compare the
source counts derived from our two independent pipelines 
and find that they agree well
within the allowance of propagated errors. In the F140W and F160W bands, the count difference
is $< 3\%$. The source magnitudes in these 
two bands are approximately 0.3 magnitude fainter
than the values measured at an aperture of 20 pixels. We correct the source
magnitude in each \HST\ band by $-0.3$ magnitude (Table 1). For the 
upper limits in other optical bands, see Fig. 4.
\vskip 0.1in

\centerline{\bf Table 1: Photometry of \emph{z} $\sim$ 9.6 Candidate in \cl}

{\noindent\small 
\begin{tabular}{ccccccccc}
\hline
\hline
F814W & F850LP & F105W & F110W & F125W & F140W & F160W & 3.6\micron & 4.5\micron \\
\hline 
$>29.4^a$ & $>27.9^a$&
$ > 28.2^a$& $27.5 \pm 0.3$ & $ 26.8\pm 0.2$ & $25.92 \pm 0.08$ & $25.70 
\pm 0.07$ & $> 26.1^a$& $24.8 \pm 0.3$ \\
\hline
\end{tabular}

$^a$  $1 \sigma$ detection limit.
}
\vskip 0.1in

\centerline{\bf 3.~~  IRAC photometry}

We retrieve archival \sst/IRAC images of \cl\
observed in July 2010 and February 2011, under Program 
ID 60034 (PI: Egami), in the form of BCD (Basic Calibrated Data) and PBCD (Post BCD). 
The BCD data are processed with tasks {\it Overlap} and {\it Mosaic} in the MOPEX package 
to produce the final mosaic images with a pixel scale of 0\asec6. Individual PBCD images serve 
as the mosaics for each of the two epochs.
The exposure times are 33.6 ksec in total and 
16.8 ksec on target for both channels at 3.6 and 4.5 $\mu$m.

As the first step, two mosaic images taken at different epochs are used.
In both epochs, the cluster is centered in the Channel 1 (3.6 \micron) and 2 (4.5 \micron) mosaics.
A visual inspection of the independent mosaics at 
4.5 \micron\ shows a clear detection at modest significance in both epochs. As the epochs are 
separated by six months, we rule out the possibility of a spurious detection or a moving object
(additional constraints in \S 8). 

The candidate 
is not visually detected at 3.6$\mu$m in either epoch, nor in the total stack,
with a $1 \sigma$ upper limit of magnitude 26.1.
The sensitivity of the $3.6 \mu$m mosaic images is estimated
by measuring the standard deviation of flux values in 2\asec4-diameter apertures 
randomly placed on empty background regions.
The IRAC photometry at 4.5\micron\ is carried out in several 
ways. We run {\it GALFIT}\cite{galfit}
to fit the brightness profile of \obj\ and neighboring sources simultaneously.
The PSF 
image is made from the 4.5$\mu$m mosaic image by stacking 
four bright (magnitude $\sim 18.5$) and isolated stars.
For bright neighboring foreground galaxies which cannot not be satisfactorily
fitted with a pure PSF model, we assume a generalized S\'{e}rsic profile and
use the higher-resolution $HST$/WFC3 $H$-band image as a reference for the
initial {\it GALFIT} input parameters.

Because of the importance of photometry in the 4.5\micron\ band, we perform
extensive tests to calibrate it. We construct simulated point sources convolved
with the IRAC PSF profile and normalized to magnitudes of 24.0, 24.5 and 25.0, respectively.
We place these simulated sources in the vicinity of \obj, 
and run {\it GALFIT} with different fitting windows (Fig.~5) until the
expected magnitude of each simulated source is recovered. We proceed to fit
the flux of \obj\ without simulated sources, using the fitting window and
background level on the image that recovered the brightness of the simulated
sources most accurately. We repeat these tests at five different positions for
the simulated sources of three different magnitudes 
to verify the measurements of
the source magnitude. 
To account for the uncertainties in estimating the background at different positions around \obj,
we choose their mean value of $24.77 \pm 0.3$ as the source magnitude in the 4.5\micron\ band.

We then use the above {\it GALFIT} results to subtract out the neighboring sources
(but not \obj\ itself), and perform aperture photometry using 2\asec4 diameter
apertures. 
The local background is determined from an annulus of radius
between 4 and 10 pixels. We find the source magnitude  $26.0 \pm 0.3$,
subject to a correction of $-0.7$ magnitude for the missing flux outside the aperture\cite{eyles}.
This magnitude derived from a small aperture is fainter by approximately
0.5 magnitude than that from {\it GALFIT} fitting. Note that the median of the pixel values in the sky
annulus is used for the background estimation. We verify that the measured source
flux could increase by $\sim 0.1-0.2$ magnitude if a smaller annulus for the sky is used
 ({\it e.g.} an outer radius of 8 pixel) to avoid the possible contamination from a few brighter
pixels. While the measurement supported the {\it GALFIT} results that the source
is detected, we do not use this value in Table 1 because of its larger uncertainty.

\vskip 0.1in

\centerline{\bf 4.~~Intermediate-redshift probability}

While our most probable and robust photometric redshift estimations yield the high-redshift solution (Fig. 3),
we further study alternative solutions when the IRAC data are not used in the fitting.
Intermediate-redshift solutions are found at low probability when
we run LPZ and BPZ with only the four WFC3/IR bands where the source is 
detected. However, these intermediate redshift solutions all have considerably
higher $\chi^2$ values than the best fit solution at $z = 9.6$. For each model used, we
calculate the $\chi^2$ value from the estimations.  
LPZ yields a best-fit model for a starburst galaxy at $z = 9.63 \pm 0.25$ 
with a low $\chi^2=0.37$.
We find a secondary solution for an elliptical galaxy at $z\sim 5.92$ with a $\chi^2= 1.74$. 
While this second solution is within $1 \sigma$ from the
best-fit value, it requires a model with an old stellar population -- an unlikely scenario at $z\sim 6$.
Fig.~6 shows the $\chi^2$ values as a function of redshifts and
the types of galaxy templates as the LPZ output. Intermediate-redshift solutions yield
considerably higher $\chi^2$ values and are hence unlikely.

\vskip 0.1in

\centerline{\bf 5.~~ Lensing model}

The basic assumption in our SL modeling approach is that light traces mass, so that the
photometry of red sequence cluster members constitutes the starting point of modeling.
We use the spectroscopic redshifts\cite{smith} and the
accurate photometric redshifts derived via the CLASH multiband imaging.
The mass model for each red sequence member is based on a surface-density power law, 
scaled by the galaxy's luminosity. The superposition of these power laws represents 
the lumpy, galaxy-scale mass component. This component is then smoothed by fitting a
low-order polynomial to it, via 2D spline interpolation,
whose result constitutes a smooth DM component.
In total there are six fundamental free parameters\cite{zitrin2}: 
the galaxy power law and the smoothing (polynomial)
degree are the first two free parameters. The two mass components are then added with a
relative galaxy-to-DM weight, which is the third free parameter.
To the resulting deflection field, we add an external shear describing the overall ellipticity.
The direction of the external shear and its magnitude are two additional
free parameters. The overall scaling of the mass model is the last free fundamental parameter.

We generate preliminary mass models expanding the six parameter space, which, along with the
CLASH imaging and photometric redshifts, help examine the multiple images and candidates
presented in previous work\cite{zitrin1,smith}  and identify two new candidate systems
(although these are not used for the minimization).
The minimization for the final best-fit models is then implemented via a Monte Carlo Markov Chain (MC) with the
Metropolis-Hastings algorithm\cite{mcmc} whose final results we use here.
The chain includes six free parameters:  the relative weight of the bright
central galaxy and five other bright galaxies in the field, which allows for a more accurate
determination of the very inner mass profile. In addition, we allow the redshift of the four
systems with photometric redshifts to vary and be optimized by the model, introducing four
additional free parameters.

The MC chain then minimizes 16 free parameters, and includes (after burn-in) a total of
20,000 steps with a typical $\sim20\%$ acceptance rate. Estimating the 
goodness-of-fit of the
best-fit model from this chain (where throughout we adopt a $\sigma=1\asec4$ as the positional
error in the $\chi^{2}$ term), the $\chi^{2}$ and root-mean-square (rms) are 61.77 and $1\asec92$, 
respectively. As constraints for the minimization, we use 23 secure multiple images of seven
sources, whereas for system 1 as defined by Zitrin \& Broadhurst\cite{zitrin1}, we use
several distinctive knots across these large images as additional
constraints. In total, we use 37 image+knots positions as constraints. The best-fit model  
yields a $\chi^2_{red} = 2.06$ and a magnification of $14.5^{+4.2}_{-1.0}$, while 
the median magnification from the MC chain is slightly higher: $15.5^{+3.3}_{-1.9}$.

In addition, we also generate a {\tt Lenstool} mass model\cite{kneib1,jullo} to compare with
our findings. The mass distribution and profile of the main halo are obtained by fitting the multiple-image
information to a Navarro-Frenk-White profile\cite{nfw}.
We include the contribution of the brightest cluster galaxy (BCG) and the 187 brightest member galaxies 
modeled by a truncated pseudo-isothermal elliptic mass distribution\cite{limousin}.   
As constraints for the minimization, we use 21 secure multiple images of seven sources, adopting a $\sigma=1\asec$ 
as the positional error.
The goodness-of-fit for this model of the 10,000 accepted samples is $\chi^{2}_{red}=2.19$, 
with an rms error of image positions of $1\asec71$ in the image plane. 
It therefore constitutes another independent measure, based on the adopted profile model that is 
different than our light-traces-mass assumption.
With this best-fit model we find a magnification of $26.6^{+20.8}_{-7.7}$ for \obj.

The magnification and shear of the two models agree within the statistical errors. However, the 
value of the magnification factor close to the critical curves is a quantity 
sensitive the model details, and one has to also examine possible systematics. 
The comparison of the two modeling methods allows us to estimate a systematic uncertainty of order of 
$\Delta\mu \sim 5$. Secondly, we check the effect of the weight (or, the mass-to-light ratio) of the bright group of 
galaxies a few arcseconds south-east of the $z\simeq9.6$ candidate image, since this could have a strong 
effect on the resulting magnification, and these are fixed in both models.
Correspondingly, we find that a reasonable $20\%$ variation in the weights of these galaxies 
entail a magnification change of $\Delta\mu\sim5$. To further examine possible 
systematics in the two
methods, we generate two independent models using {\tt Lenstool} (the first includes one central DM component, 
while the other which we incorporate here, models the cluster as two DM clumps), and several independent 
models using the  method of Zitrin et al. (2011)\cite{zitrin2}, each with a different combination of free parameters ({\it i.e.}, 
different galaxies freely weighted, or photometric redshift optimizations). By doing so, we have  a set 
of models to compare to the best fitting model used above, and assess the systematics these changes 
entail. We note that some of these resulting models have critical curves (where the magnification diverges) 
that pass through the image or further outwards of its location, so that in principle, the upper systematic 
limit on the best-fit model magnification, is poorly constrained. Both of these
models 
have in general a lower reduced $\chi^{2}$ than the complementary chain models,
and in particular, yield better reproduction of images, as gauged by visual inspection. These are therefore chosen as the 
best-fit models used above whose values we adopt throughout. We conclude that systematic uncertainties are 
of the same order as the statistical uncertainties, albeit with poorly constrained upper limits.

Our best fit lens model predicts that \obj\ lies outside the caustics for $z\sim10$ in the
source plane, so that it likely not to be multiply-lensed, but it is still
predicted to be significantly magnified by gravitational lensing from the cluster. 
To examine the possibility of multiple images, we choose a complementary MC chain
model with somewhat a different combination of free parameters that does predict multiple images.
No counter images brighter than magnitude 27 are found in the area where counterpart images are predicted.

Both lensing models predict a highly elongated image. Although the observed 
source is elongated along in the direction predicted by the lensing model,
the level of elongated is lower than the model predictions. 
As a result, the intrinsic image in the source plane is not circular. 
However, as the magnification especially close to the critical curves is one 
of the more sensitive quantities to measure, this probably results from 
statistical and potential systematic uncertainties in lens modeling. 
We anticipate future improvements in modeling to reduce such potential 
systematic errors. 

\vskip 0.01in

\centerline{\bf 6.~~ Image Deconvolution}

The drizzled \H-band image is deconvolved using 20 iterations of
Lucy-Richardson deconvolu\-tion\cite{lucy,richardson}.
The PSF is provided by a bright field star within the
stacked image displaced $39\arcsec$ from the object. 
This is a modest amount of deconvolution, but is sufficient to remove 
the blurring due to the PSF wings, and to provide a
model-independent representation of the object. 

After deconvolution, the extent of the source is significant out to
$\approx 0\asec3$ from its center. The distribution of light
for $0\asec13\leq r \leq0\asec33$ is roughly exponential
with a scale length of $0\asec067\pm0\asec005$. The isophote ellipticity
over the region fitted increases with radius to $\sim0.5.$ at $r=0\asec33$. 
The half-light radius of the source is $r<0\asec13.$ 

Using the  detailed lens model for \cl\ and the PSF-deconvolved F160W
image, we reconstruct the \obj\ image in the source plane at $z=9.6$.
The source-plane reconstruction is shown in Fig.~7. The candidate
is significantly elongated in the source plane along a position angle
of 139 degrees and is well-fit by a 2D Gaussian.  The Gaussian fit shows
that the candidate has an axis ratio of 7.55.  Assuming 4.28 kpc
arcsec$^{-1}$ at $z=9.6$, the source spans 1.28 kpc and 0.17 kpc along its
major and minor axis, respectively. Note that the most significant elongation 
is in the F140W band.

\vskip 0.1in

\centerline{\bf 7.~~ Spectral Energy Distribution Modeling} 

Our SED modeling constructs a large suite of models using Monte Carlo
draws of the free parameters, and then evaluates the posterior
probability distribution of each parameter by calculating the
statistical likelihood of each model.  To fit \obj{} we synthesize
photometry in the 19 observed bands for $50,000$ models assuming a
fixed redshift of $z=9.6$.  We parameterize the star formation history
$\psi(t)$ as a delayed $\tau$ model, $\psi(t)\propto t\exp(-t/\tau)$,
where $t$ is the time since the onset of star formation and $\tau$ is
a characteristic time scale.  The advantage of this parameterization
is that it allows for both linearly rising ($t/\tau\ll1$) and
exponentially declining ($t/\tau\gg1$) star formation histories.  We
draw $\tau$ from a uniform distribution of between 10~Myr and 1~Gyr, 
$t$ uniformly from $5-500$~Myr, and the stellar metallicity in the range
$Z=0.002-0.02$ ($10\%-100\%$ solar).  For our fiducial modeling we
assume no dust obscuration, although we test the effect of relaxing
this assumption below.

Fig.~8 shows the posterior probability distributions on the stellar
mass, star-formation rate (SFR), and SFR-weighted age, $\langle
t\rangle_{\rm SFR}\equiv \int_{0}^{t}
\psi(t^{\prime})(t-t^{\prime}) \,{\mathrm d}t^{\prime} / \int_{0}^{t}                                  
\psi(t^{\prime}) \,{\mathrm d}t^{\prime}$.  
Since we are unable to place any significant constraints on
either $\tau$ or the metallicity $Z$, 
we do not show these
probability distributions here.  Based on the median of the posterior
probability distributions, our Bayesian analysis suggests a stellar
mass of $\sim 1.5 \times 10^8~(\mu/15)^{-1}$ \Msun\ for
\obj, and a SFR of $\sim 1.2~(\mu/15)^{-1}$~\Msun~yr$^{-1}$.  Although
the probability distribution on $\langle t\rangle_{\rm SFR}$ is not
peaked, we find that $95\%$ of the models have $\langle t\rangle_{\rm
SFR}<200$~Myr, suggesting a likely formation redshift $z_{f}<14.2$.
This analysis clearly demonstrates the need for precise IRAC
photometry of high-redshift galaxies, in order to place their
physical properties on firmer quantitative footing.

We investigate the effect of changing our prior assumptions on
these results. First, we consider an ensemble of models that
includes dust, allowing the rest-frame $V$-band
attenuation\cite{calzetti} to range from $0-2$~magnitude.  This
analysis yield median stellar mass and SFR estimates that are a
factor of $\sim6$ higher, but with no improvement in the likelihood,
and no constraint on the $V$-band attenuation.  However, our
constraints on $z_{f}$ change by $<5\%$.  We also consider simple
exponentially declining star formation histories; those yielded
similar estimates for the stellar mass and SFR (within a factor of
$\sim2$) with respect to our fiducial model parameters, and
constrain the formation redshift to $z_{f}<17.5$.

We also test the possibility that \obj{} is an intermediate-redshift
interloper (see Fig.~9). Assuming a representative value of $z=3.2$
(Fig.~3 and 6), we construct a suite of $50,000$ models with exponentially declining
star formation histories spanning a wide range of stellar metallicity
($0.0002-0.03$), dust attenuation ($A_{V}=0-3$~magnitude), age ($5$~Myr to
$2$~Gyr , where $2$~Gyr is the age of the Universe at this redshift),
and $\tau$ ($0.01-10$~Gyr).  Assuming an intermediate-redshift solution, the
best-fit model has a reduced $\chi^{2}=4.1$, compared to
$\chi^{2}=1.4$ when assuming a redshift $z=9.6$.  In addition, as
shown in the inset to Fig.~9, the $\chi^{2}$ distribution of the
full suite of models assuming $z=9.6$ all peak around
$\chi^{2}\approx1.5$ (light blue histogram), whereas the $\chi^{2}$
distribution of the bulk of the models assuming $z=3.2$ are centered
around $\chi^{2}\approx9$ (gray histogram).  This analysis
demonstrates that the high-redshift solution is clearly preferred from
the point-of-view of the SED modeling.

The age estimate of high-redshift galaxies is useful, as magnified
sources make it significantly easier to carry out
\sst/IRAC photometry. Without the lensing effect, the rest-frame
optical flux would be virtually impossible to measure for galaxies at
$z > 8$, as the anticipated fluxes in the IRAC bands are below the
IRAC confusion limits.  
Fig.~10 plots the \HST\ and IRAC magnitudes for a number of objects\cite{labbe2} 
at $z \sim 7-8$. Our measurements are consistent with the result, but with
considerably higher accuracy at the source's intrinsic magnitude, thanks to the 
gravitational magnification.

\vskip 0.1in

\centerline{\bf 8.~~ Solar System and Galactic Interlopers}

We demonstrate here that the likelihood that \obj\ is either a faint solar system 
or Galactic object is extremely low. CLASH observations of \cl\ were obtained at 
eight different epochs and our F140W and F160W images, in particular, cover
five of those
epochs spanning 80 days between December 20, 2010 and March 10, 2011. At each 
epoch, we measure the relative separation between \obj\ and a bright 
early type galaxy with a compact core that is $\sim 10\arcsec$ away. This 
galaxy provides our common stationary astrometric reference point. The relative 
offsets, as a function of time, with respect to the mean separation between \obj\ 
and the reference galaxy are shown in Fig.~11. Based on these 
measurements, the proper motion of the source is $<0\asec13$ per year. If 
\obj\ were an object on a low-eccentricity orbit within the solar system, its orbital 
period would have to be in excess of 10 million years - implying an orbital 
semi-major axis that is at least two orders of magnitude beyond the distance of 
Kuiper belt and Trans-Neptunian objects ($40 - 100$ AU). Only objects in the Oort 
cloud ($\sim 50,000$ AU) would be expected to have such small proper motions but 
both the predicted absolute magnitudes and colors of typical Oort cloud 
objects\cite{oort} 
would be inconsistent with those of \obj\ (\obj\ is about 9 magnitudes brighter 
than what is expected for a 20 km wide Oort cloud object).

To assess whether \obj\ could be a cool Galactic star, we compare its colors to a 
sample of 75 stellar templates of L,M,T dwarfs compiled from a spectral 
atlas\cite{rayner,cushing} and Y dwarfs from stellar model 
atmospheres\cite{huberny}. For each star we
compute its predicted flux in F814W, F105W, F110W, F125W, F140W, F160W, IRAC 3.6 \micron\ and
4.5 \micron\ bands and normalize the fluxes so they match the F160W measurement 
for \obj. We then compute, for each star, the corresponding total flux deviation 
from the \obj\ flux values using the expression:
\begin{equation}
\left< \Delta f \right> = \sqrt{\Sigma_{i=1}^{N} \left[ (f_{i,\obj} - f_{i,STAR})/\sigma_{i,obj}\right]^2}
\end{equation}
where the sum is over the eight bands. 
There are no L,M,T, or 
Y dwarfs whose total 
flux difference is within $6 \sigma$ of the observed colors of \obj (see also Fig.~12). Indeed, 
the minimum difference in flux space between the source and the closest stellar match is $6.8 \sigma$ 
and the median difference is $16.8 \sigma$.
The combination of NIR detections and upper limits measured for \obj\ thus argues 
strongly against a cool, faint Galactic star as the likely explanation for the source. 
\vskip 0.1in

\centerline{\bf 9.~~ Photometric Scatter Test}

We demonstrate that the photometry of \obj\ is not likely to be due to 
drawing randomly from the main faint galaxy population. We extract from all 12 
CLASH cluster object catalogs derived from IR-based detection images those sources 
that have approximately similar WFC3 F160W fluxes as \obj. The selection 
criterion is $25.45 < {\rm F160W} \le 26.95$. There are a total of 5614 objects 
that satisfy this criterion. We then generate new magnitudes for each object by 
randomly drawing, from a Gaussian distribution, with a mean equal to the object's 
measured flux and with a standard deviation equal to the object's flux uncertainty.
 We then count how many such objects would have measured flux ratios that lie 
within the 1$\sigma$ uncertainty of the F110W/F125W and F125W/F160W flux ratios of
\obj\ and that also show no flux (at the 2$\sigma$ level) in the F814W and F105W 
bands. We run 1000 such realizations of the sample and find that only 0.015\% of 
the objects (averaged over 1000 realizations) would satisfy these criteria. Thus, 
\obj\ is unique amongst the population of faint sources at the 99.985\% level
 ( Fig.~13).

\vskip 0.1in

\centerline{\bf 10.~~ Summary}

We carry out extensive analyses to study the nature of such an object with unique properties 
and found the following evidence:

\begin{itemize}
\item[$\bullet$] Detection of the source at multiple epochs in both \HST\ and \sst\ imaging rules
 out the source being a spurious detection or a transient object.
\item[$\bullet$] The combination of color decrements at $\sim 1.3$ \micron\ and $\sim 4$
 \micron\ favor a high-redshift solution
\item[$\bullet$] Intermediate-redshift fits yield significantly higher values of 
 $\chi^2$ using just \HST\ photometry alone.
\item[$\bullet$] Intermediate-redshift solutions are ruled out at $4 \sigma$ when
 \HST\ and \sst\ photometry are used in photometric redshift estimation. 
\item[$\bullet$] The de-lensed magnitude is consistent with expectations for sources at $z > 8$.
\item The de-lensed half-light radius is more consistent with expectations for sources at $z > 8$.
\item[$\bullet$] The source plane morphology is significantly more elongated 
 than the image plane morphology. 
\item[$\bullet$] The proper motion upper limit and the source's apparent magnitude make it very
 unlikely that the source is a Kuiper Belt, Trans Neptunian or Oort cloud object.
\item[$\bullet$] Location of the source in multi-color space is inconsistent (at $4 \sigma$)
 with the source being a cool Galactic dwarf star (spectral type L,M,T,Y).
\item[$\bullet$] Photometric scatter is not sufficient to explain colors, and this explanation 
 is rejected at the 99.985\% confidence level.
\end{itemize}

We therefore conclude that the \obj\ is very highly unlikely to be an intermediate-redshift interloper, 
a cool, late-type star, or solar system object. The most probable explanation 
would seem to be a $z = 9.6$ galaxy.

\clearpage

\begin{figure} 
\parbox{5.5 in}{
\includegraphics{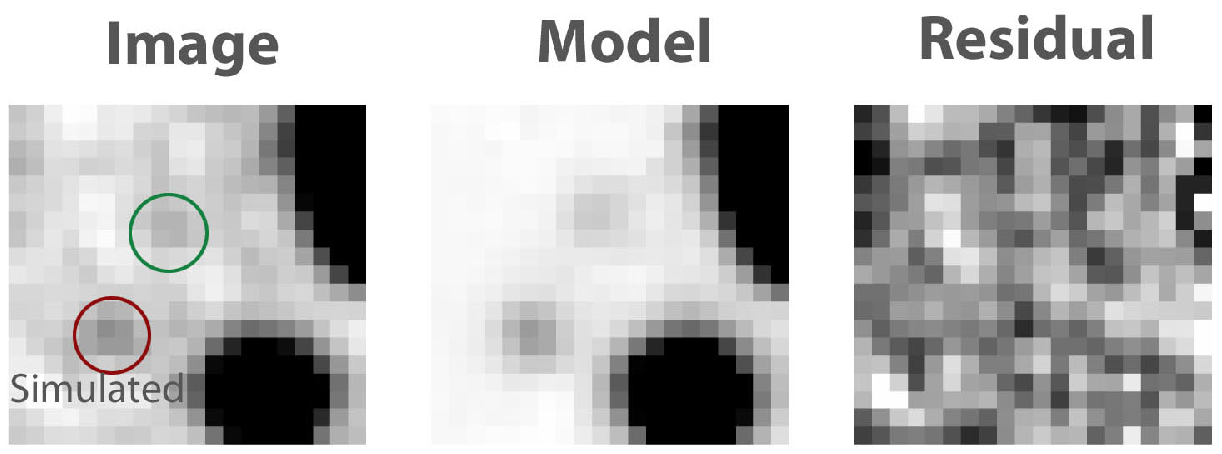} 
}

\vspace{5.95 in}
\noindent {\sf Figure~5 --  
Illustration of IRAC fitting at 4.5\micron. In the left panel, 
\obj\ is marked with a green circle, and a simulated point
source of AB=24.0 is marked with a red circle. In the middle panel, the
best-fit {\it GALFIT} model is displayed, and in the right panel,
the residual image with all model components subtracted. Note that the actual
fitting is made without simulated sources and yields a mean magnitude of 
24.8$\pm$0.3.
}
\label{galfit}
\end{figure}
\clearpage

\begin{figure} 
\parbox{5.5 in}{
\includegraphics{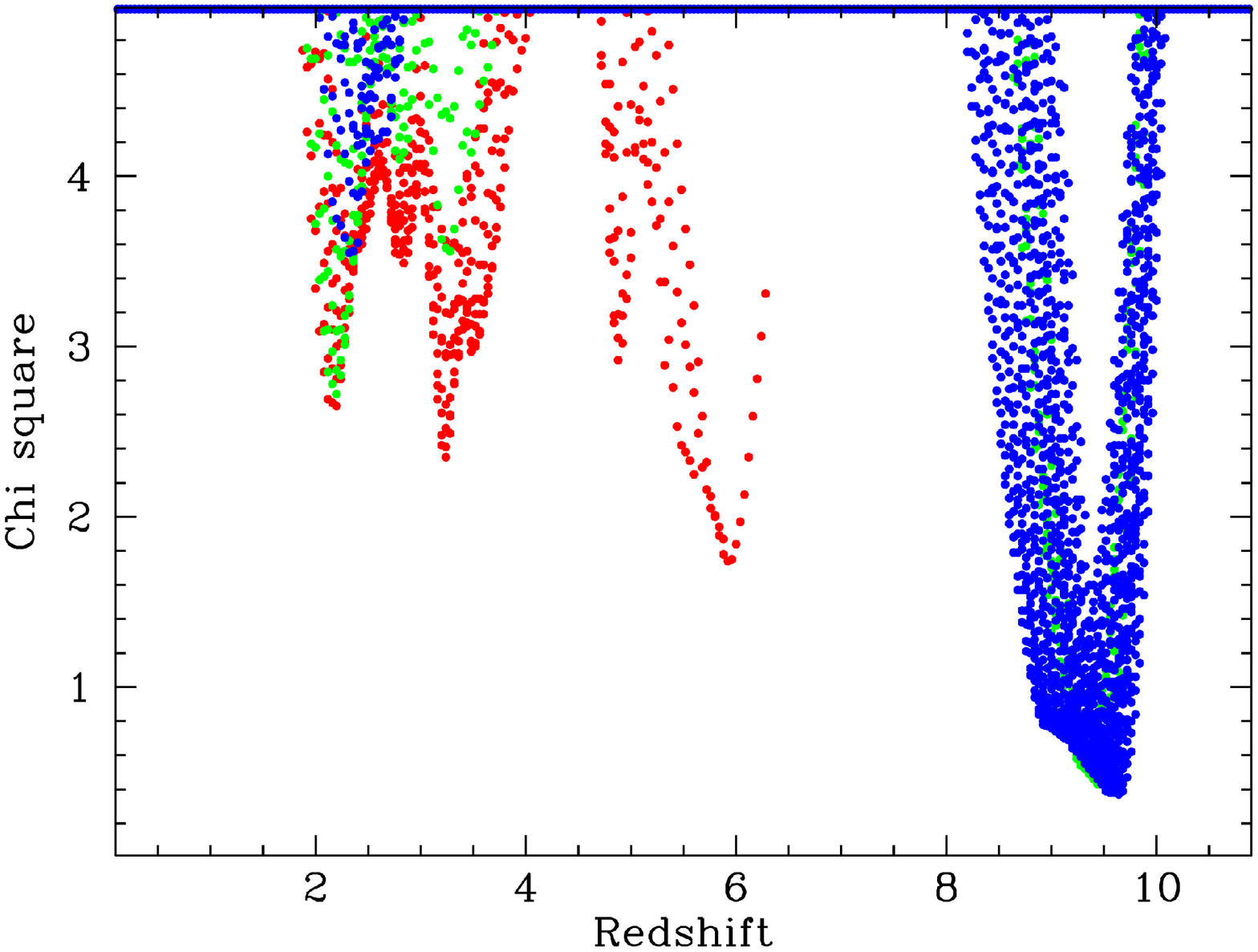} 
}

\vspace{5.95 in}
\noindent {\sf Figure~6 --  
Likelihood distribution of photometric redshift. We fit only the four \HST\ bands where the source 
is detected, and then plot the $\chi^2$ values at different fitted redshifts for each template in
the LPZ template library, plus the effect of dust attenuation. 
$\chi^2$ values higher than 5 are truncated. Green points: elliptical galaxies; red points:
spiral galaxies; and blue points: starburst galaxies, some of which yield the lowest $\chi^2$ values. 
Intermediate-redshift solutions yield considerably higher $\chi^2$ values than the high-redshift solutions.
}
\end{figure}
\clearpage 

\begin{figure} 
\parbox{5.5 in}{
\includegraphics{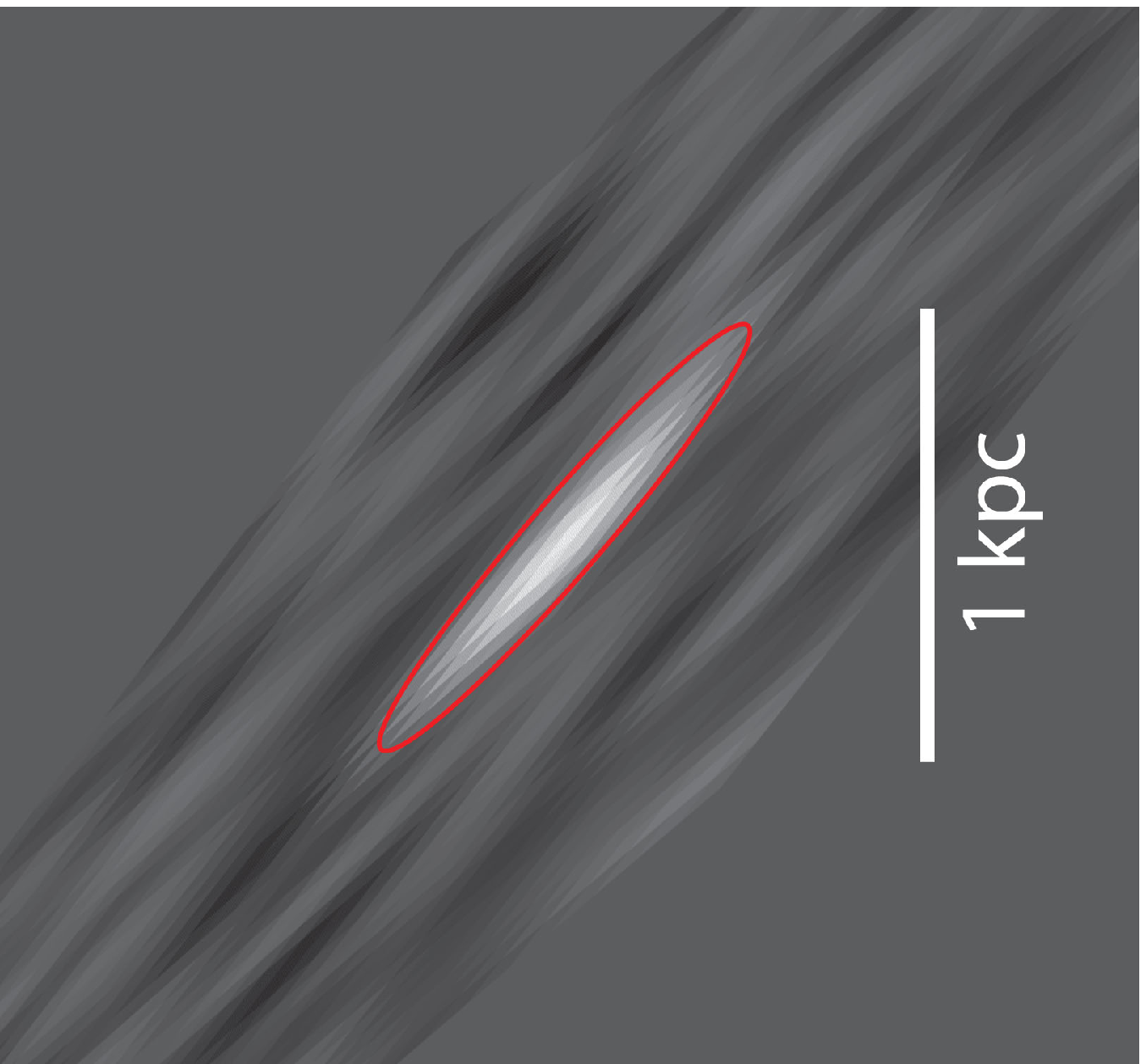} 
}

\vspace{5.95 in}
\noindent {\sf Figure~7 --  
Source-plane reconstruction of \obj\ in the WFC3/IR F160W band.  The
source is significantly elongated with an axis ratio of 7.55.  The
candidate spans 1.28 kpc and 0.17 kpc along its major and minor axis,
respectively, as denoted by the red ellipse. The results are sensitive to 
the model details and systematic errors. 
}
\label{deconv}
\end{figure}
\clearpage 

\begin{figure} 
\parbox{5.5 in}{
\includegraphics{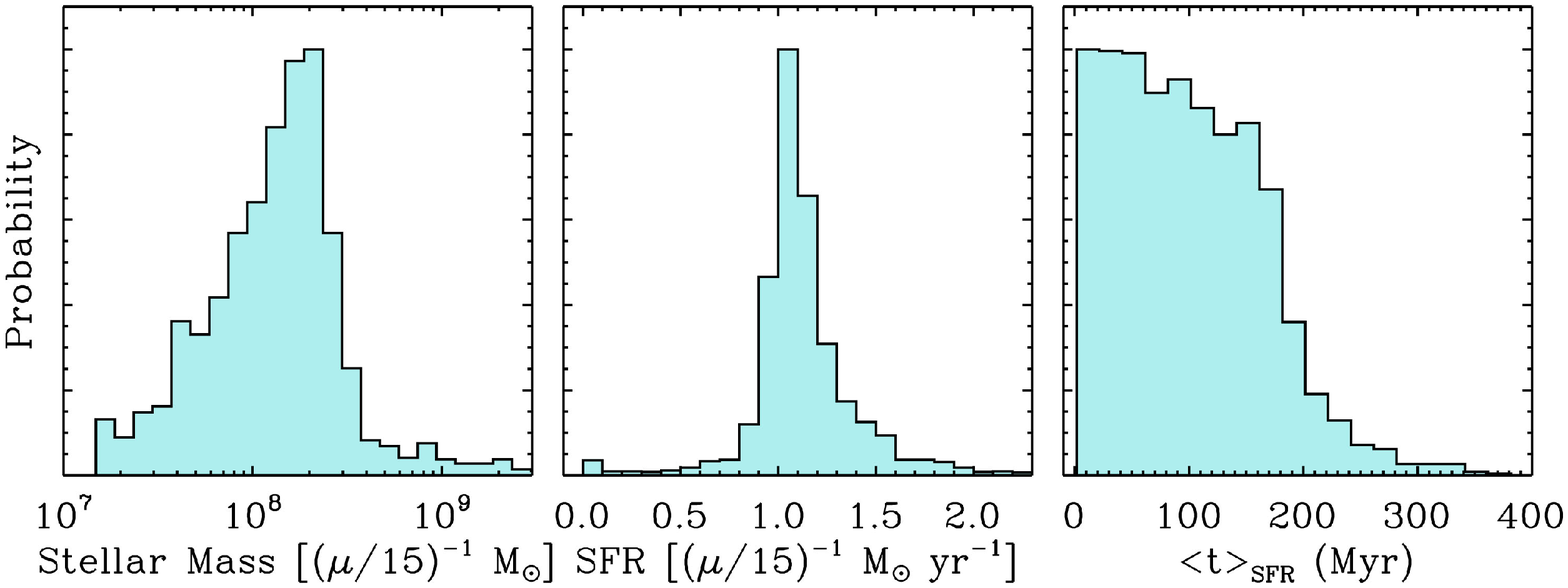} 
}

\vspace{5.95 in}
\noindent {\sf Figure~8 --  
Posterior probability distributions on the stellar mass, SFR, and
SFR-weighted age, $\langle t\rangle_{\rm SFR}$ based on our Bayesian
SED modeling.  Note that our stellar mass and SFR estimates have been
de-magnified assuming a fiducial magnification factor $\mu$=15, while
$\langle t\rangle_{\rm SFR}$ is independent of $\mu$.  Based on this
analysis we infer a stellar mass of $\sim
1.5\times10^8~(\mu/15)^{-1}$ \Msun{} a star-formation rate (SFR) of
$\sim1.2~(\mu/15)^{-1}$~\Msun~yr$^{-1}$, and a constrain on the
SFR-weighted age of $<200$~Myr ($95\%$ confidence level), implying a
formation redshift $z_{f}<$14.2.}
\label{posteriors}
\end{figure}
\clearpage 

\begin{figure} 
\parbox{5.5 in}{
\includegraphics{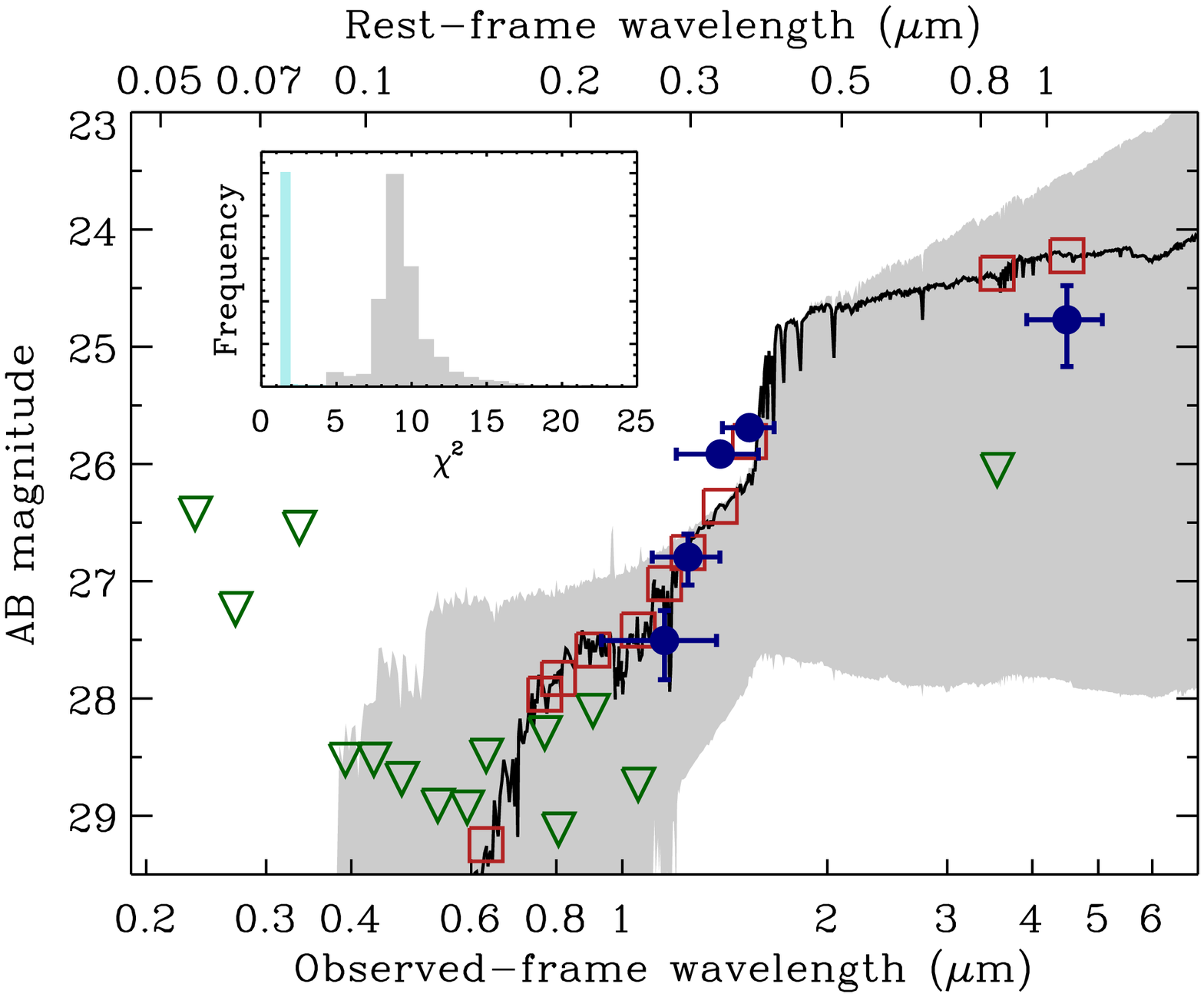} 
}

\vspace{5.95 in}
\noindent {\sf Figure~9 --  
Results of modeling the SED of \obj{} assuming an intermediate-redshift
solution, \emph{z}=3.2.  This figure is analogous to Fig.~4 but here the
gray shading shows the range models drawn from the posterior
probability distribution that fit the data assuming \emph{z}=3.2.  As shown
in the inset, the $\chi^{2}$ distribution of these intermediate-redshift models
peaks around $\chi^{2}\approx$9 (gray histogram), whereas the
$\chi^{2}$ distribution of the models fitted to the data assuming
\emph{z}=9.6 peaks around $\chi^{2}\approx$1.5 (light blue histogram; see
also Fig.~4).  We conclude, therefore, that the high-redshift solution
is clearly preferred from the point-of-view of our SED modeling.}
\label{lowz_sed} 
\end{figure}
\clearpage

\begin{figure} 
\parbox{5.5 in}{
\includegraphics{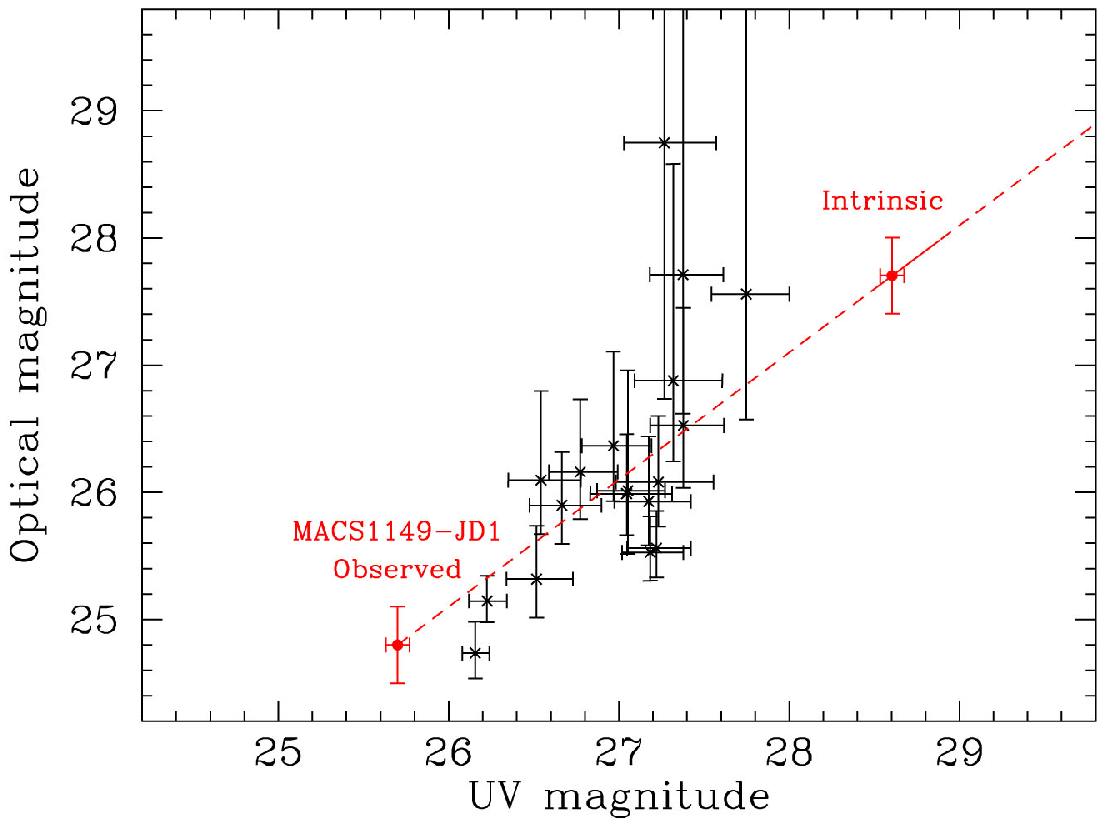} 
}

\vspace{5.95 in}
\noindent {\sf Figure~10 --  
Rest-frame UV and optical magnitudes of high-redshift objects. 
The data\cite{labbe2} are from objects at \emph{z}$\sim$7-8.
The rest-frame UV band is F125W, and the rest-frame optical band
is the IRAC 3.6\micron. The magnitudes
of \obj, in F160W and 4.5 \micron\ bands, are plotted in red,
in their observed values (the lower point) and de-magnified values (scaled down by a
flux factor of 15, at the upper-right).
The dashed red line is the track of the source's intrinsic magnitudes under 
different magnification factors, and the solid red line marks the range for \obj.
Cluster lensing makes it possible to improve the accuracy of photometry at the faint end.
}
\label{labbe}
\end{figure}
\clearpage

\begin{figure} 
\parbox{5.5 in}{
\includegraphics{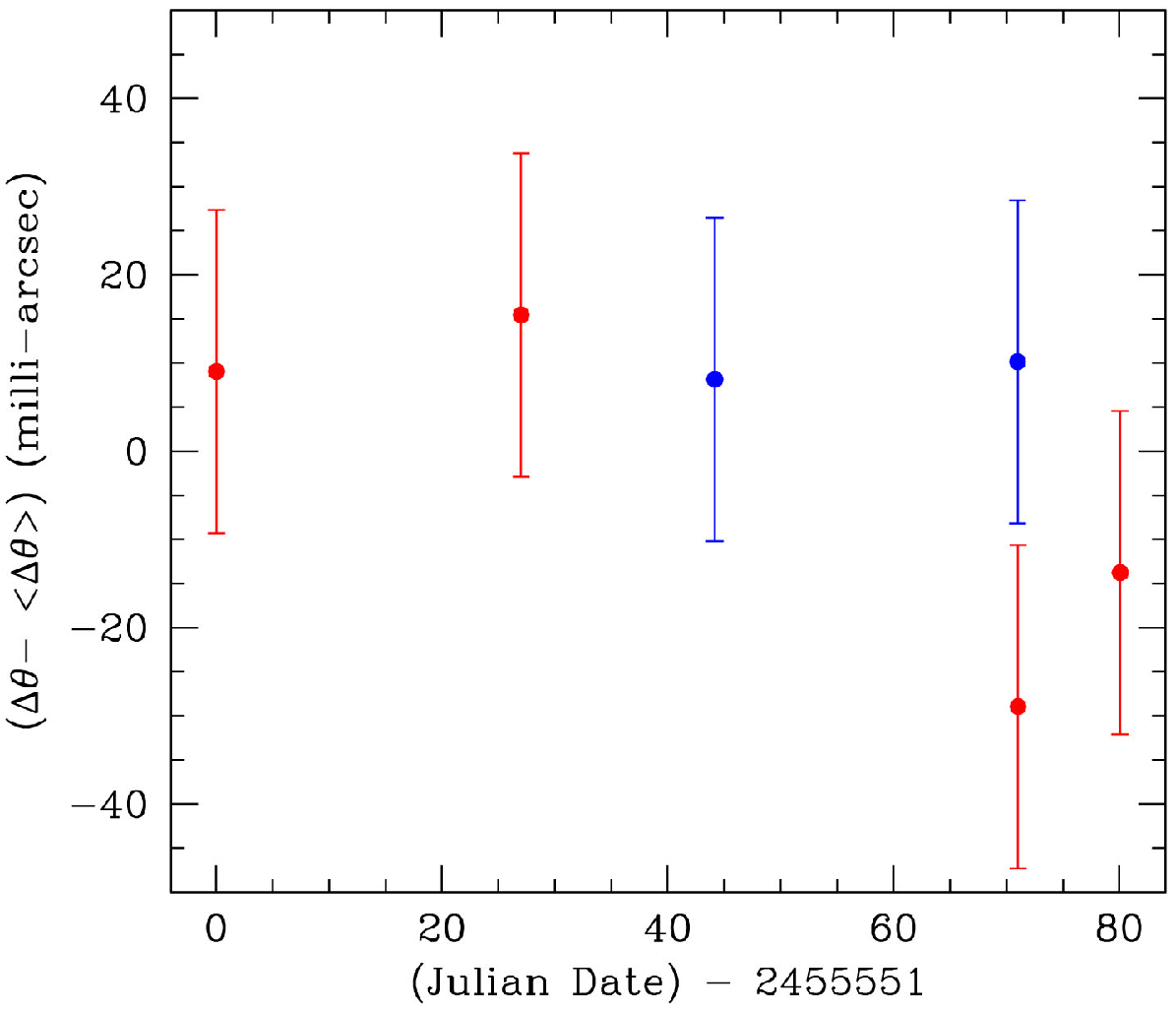} 
}

\vspace{5.95 in}
\noindent {\sf Figure~11 --  
Relative difference, in milli-arcseconds, between the separation of \obj\ from 
a nearby reference galaxy at each of five separate epochs and the mean 
separation value. The upper limit on its proper motion is $< 0\asec13$ yr$^{-1}$.
}
\label{centroid}
\end{figure}
\clearpage

\begin{figure} 
\parbox{5.5 in}{
\includegraphics{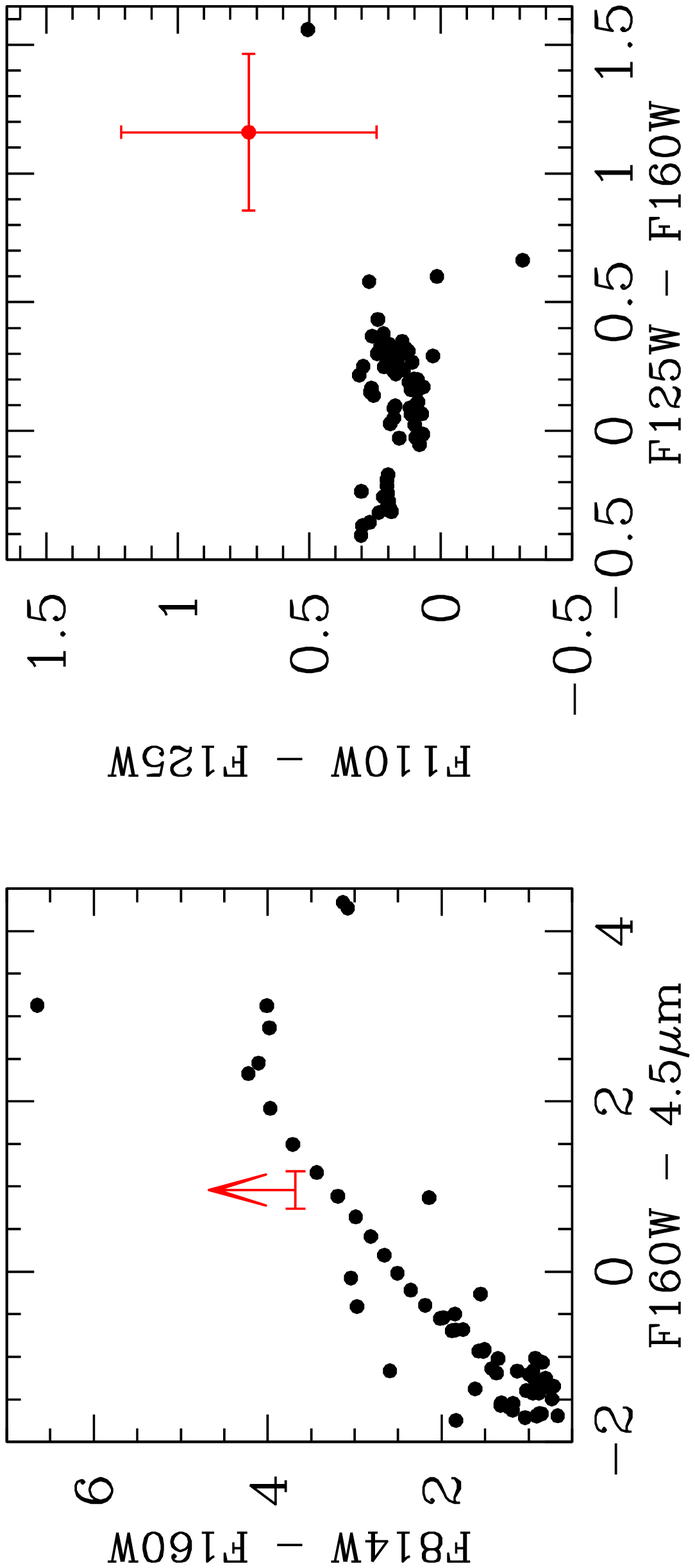} 
}

\vspace{5.95 in}
\noindent {\sf Figure~12 --  Comparison of colors of 75 late-type 
stars\cite{rayner,cushing,huberny} (black circles) and those of the high-redshift 
candidate \obj\ (red circle).
The F814W magnitude for \obj\ is based on its $1 \sigma$ upper limit.
In the right panel, the color of a rare M-8III star is close to the error box,
but it is well separated in color in the left panel. 
}
\label{dwarf}
\end{figure}
\clearpage

\begin{figure} 
\parbox{5.5 in}{
\includegraphics{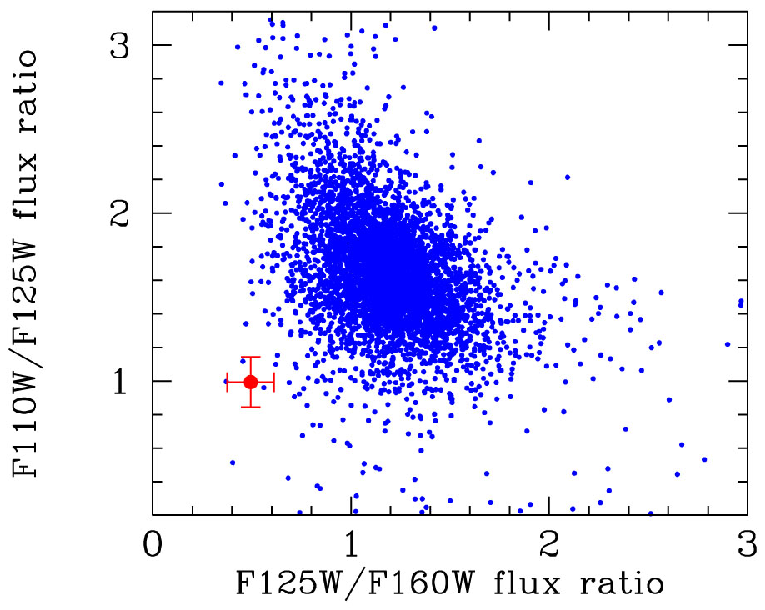} 
}

\vspace{5.95 in}
\noindent {\sf Figure~13 --
Flux ratios of faint galaxy population
in the CLASH database with F160W magnitude 25.45$-$26.85. 
\obj, a resolved source is marked by a red circle. 
The plotted fluxes are isophotal values and hence are slightly different from Table 1.
The five sources next to or within the error box are further examined, and they are
rejected because of a detection in optical bands.
}
\label{scatter}
\end{figure}
\clearpage

\end{document}